\documentclass[preprint,showpacs,preprintnumbers,amsmath,amssymb,pra,superscriptaddress]{revtex4}

\usepackage[dvips]{graphicx} 
\usepackage{amsfonts}
\usepackage{amscd}
\usepackage{amsmath}    

\usepackage{subfigure}

\begin{document}
\title{Experimental demonstration of phase-remapping attack in a practical quantum key distribution system}

\author{Feihu Xu}
 \email{feihu.xu@utoronto.ca}

 \author{Bing Qi}
  \email{bqi@physics.utoronto.ca}

 \author{Hoi-Kwong Lo}
  \email{hklo@comm.utoronto.ca}

\affiliation{ Center for Quantum Information and Quantum Control
(CQIQC),
Dept. of Physics and Dept. of Electrical and Computer Engineering,\\
University of Toronto, Toronto, Ontario, M5S 3G4, Canada }

\begin{abstract}
Unconditional security proofs of various quantum key distribution
(QKD) protocols are built on idealized assumptions. One key
assumption is: the sender (Alice) can prepare the required quantum
states without errors. However, such an assumption may be violated
in a practical QKD system. In this paper, we experimentally
demonstrate a technically feasible ``intercept-and-resend" attack
that exploits such a security loophole in a commercial ``plug \&
play" QKD system. The resulting quantum bit error rate is 19.7\%,
which is below the proven secure bound of 20.0\% for the BB84
protocol. The attack we utilize is the phase-remapping attack (C.-H.
F. Fung, et al., Phys. Rev. A, 75, 32314, 2007) proposed by our
group.

\end{abstract}


\maketitle

\section{Introduction}
Quantum key distribution (QKD) \cite{quantum2, quantum, quantum1}
enables an ultimately secure means of distributing secret keys
between two parties, the sender (Alice) and the receiver (Bob). In
principle, any eavesdropping attempt by a third party (Eve) will
unavoidably introduce disturbance and be detected. The unconditional
security of the BB84 QKD protocol has been rigorously proved based
on the laws of quantum mechanics \cite{security proof}, even when
implemented on imperfect practical setups \cite{gllp}.

Unfortunately, a practical QKD system has imperfections. Eve may try
to exploit these imperfections and launch specific attacks not
covered by the original security proofs. Is it possible that a small
unnoticed imperfection spoils the security of the otherwise
carefully designed QKD system? This question has drawn a lot of
attention. Gisin et al. studied a Trojan-horse attack employing the
unwanted internal reflection from a phase modulator \cite{Trojan}.
Makarov et al. proposed a faked state attack, which exploits the
efficiency mismatch of two detectors in a practical QKD system
\cite{fake}. Our group has proposed \cite{time shift} a simple
attack---time-shift attack---that exploits the same imperfection.
Moreover, we have experimentally demonstrated \cite{time shift exp}
our attack against a commercial QKD system. This was the first time
that a commercial QKD system had been successfully hacked, thus
highlighting the vulnerability of even well-designed commercial QKD
systems. Lamas-Linares et al. demonstrated that the information
leakage due to public announcement of the timing information can be
used by Eve to access part of the sifted keys \cite{leak}. Recently,
the imperfections of some particular single photon detectors, namely
those based on passively or actively quenched avalanche photodiodes,
and potential security loopholes of practical QKD systems employing
such detectors have been studied in \cite{blind}. For more general
discussions, see \cite{attack}.

A key assumption in QKD is Alice encodes her signals correctly. This
seems like a simple assumption. However, it may be violated in
practice by, for example the phase-remapping attack
\cite{remapping}. In this paper, we experimentally investigate the
phase-remapping attack in a commercial QKD system. In contrast with
the time-shift attack, the phase-remapping attack is an
``intercept-and-resend" attack which allows Eve to gain the full
information of the sifted keys. Here, we experimentally find that
the phase remapping process in a practical QKD system is much more
complicated than the theoretical model in Ref. \cite{remapping}. To
adapt to this complexity, we have modified the original
phase-remapping attack into type 1 and type 2 practical attacks,
where type 2 attack is more practical in QKD systems. It is well
known that in the standard BB84 QKD system, a simple
``intercept-and-resend" attack will introduce a quantum bit error
rate (QBER) of 25\% which alarms the users that no secure keys can
be generated. Our experimental results show that by performing the
phase-remapping attack, Eve can gain the full information at the
cost of introducing a QBER of 19.7\%, which is lower than the proven
security bound of 20.0\% for the BB84 protocol \cite{bb84 proof}. In
other words, the security of the commercial QKD system is
compromised.

This paper is organized as follows: in Section \ref{theory}, we
summarize the basic idea of phase-remapping attack and then propose
our theoretical model to analyze QBER. In Section \ref{Exp}, we
describe the phase modulation scheme adopted in a commercial ``plug
and play" QKD system and discuss two types of practical attacks
which could be applied to this specific design. In Section
\ref{result}, we show the experimental results and analyze the QBER.
We finally conclude our paper with some general comments in Section
\ref{con}.

\section{Phase-remapping attack} \label{theory}
Practical limitations associated with phase and polarization
instabilities over long distance fibers have led to the development
of bidirectional QKD schemes, such as the ``plug-and-play'' QKD
structure \cite{plug and play} and the Sagnac QKD structure
\cite{Sagnac}. Specially, the ``plug-and-play'' structure is widely
used in commercial QKD systems \cite{commercial}. In this system,
Bob first sends two strong laser pulses (signal pulse and reference
pulse) to Alice. Alice uses the reference pulse as a synchronization
signal to activate her phase modulator. Then Alice modulates the
phase of the signal pulse only, attenuates the two pulses to single
photon level, and sends them back to Bob. Bob randomly chooses his
measurement basis by modulating the phase of the returning reference
pulse. Since Alice allows signals to go in and go out of her device,
this opens a potential back door for Eve to launch various attacks
\cite{Trojan}. One specific attack is the phase-remapping attack
\cite{remapping}.

\begin{figure}[!t]
\resizebox{7cm}{!}{\includegraphics{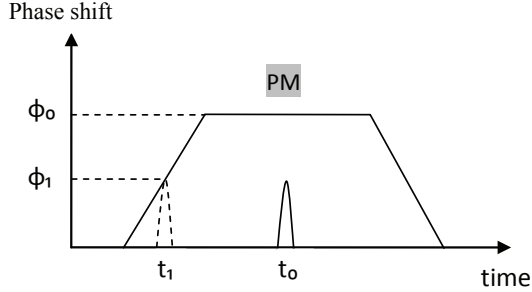}} \caption{Diagram
of phase modulation (PM) signal. $t_{0}$ is the original time
location where Bob's signal pulse is properly modulated to have
phase $\phi_{0}$. Eve time shifts the signal pulse from $t_{0}$ to
$t_{1}$. This pulse will undergo a new modulated phase $\phi_{1}$.}
\label{Fig.1}
\end{figure}

LiNbO$_{3}$ waveguide phase modulator is commonly used to encode
random bits in fiber based phase-coding BB84 QKD system. In
practice, a phase modulator has finite response time, as shown in
Fig. \ref{Fig.1}. Ideally, Bob's signal pulse passes through Alice's
phase modulator in the middle of the modulation signal and undergoes
a proper modulation (time $t_{0}$ in Fig. \ref{Fig.1}). However, if
Eve changes the time difference between the reference and the signal
pulse, the signal pulse will pass through the phase modulator at a
different time (time $t_{1}$ in Fig. \ref{Fig.1}), and the encoded
phase will be different. Originally, Alice uses \{0, $\pi/2$, $\pi$,
$3\pi/2$\} to encode \{$0_{1}$(bit ``0" in base1), $0_{2}$(bit ``0"
in base2), $1_{1}$(bit ``1" in base1), $1_{2}$(bit ``1" in base2)\}.
Now, after Eve's remapping process, Alice's encoded phases \{0,
$\pi/2$, $\pi$, $3\pi/2$\} will be mapped to \{0, $\phi_{1}$,
$\phi_{1}+\phi_{2}$, $\phi_{1}+\phi_{2}+\phi_{3}$\}, where
$\phi_{i}$ (i=1,2,3) is the new phase difference between two
adjacent states. Ref \cite{remapping} assumed for simplicity, that
the phase modulator has the same rising time and proportional phase
modulation for each encoded phase, i.e.
$\phi_{1}$=$\phi_{2}$=$\phi_{3}$. Here, we consider a more general
setting. The exact value of $\phi_{i}$ depends on time displacement
introduced by Eve and the actual phase modulation system. This phase
remapping process allows Eve to launch a novel
``intercept-and-resend" attack: phase-remapping attack. In our
experiment, the practical attack strategy is:

(1) Eve intercepts Bob's strong pulse and sends a time-shifted pulse
to Alice via her own device. Note that Eve can change the actual
values of $\phi_{i}$(i=1,2,3) by changing the time displacement.
However, she cannot change $\phi_{1}$, $\phi_{2}$, and $\phi_{3}$
independently.

(2) Eve's strategy is to either distinguish \{$0_{1}$\} from
\{$0_{2}$, $1_{1}$, $1_{2}$\} or \{$1_{2}$\} from \{$0_{1}$,
$0_{2}$, $1_{1}$\} with minimal errors. To distinguish \{$0_{1}$\},
Eve introduces a phase shift of \{$\phi_{1}+\phi_{2}$\} by using her
phase modulator on the reference pulse sent back by Alice and
performs an interference measurement. If detector1 has a click
\cite{det1}, Eve sends a standard BB84 state \{$0_{1}$\} to Bob.
Otherwise, Eve simply discards it. A similar procedure is performed
to distinguish \{$1_{2}$\}, where Eve introduces a phase shift of
\{$\phi_{1}$\} on the reference pulse. Here, we define Eve's phase
shift \{$\phi_{1}$\} as Base1, \{$\phi_{1}+\phi_{2}$\} as Base2.

Now, assume that Eve uses Base2 to distinguish \{$0_{1}$\}; given
Alice sends different states \{$0_{1}$, $0_{2}$, $1_{1}$, $1_{2}$\},
Det1's detecting probabilities \{$P_{0_{1}}$, $P_{0_{2}}$,
$P_{1_{1}}$, $P_{1_{2}}$\} are
\{$\sin^{2}(\frac{\phi_{1}+\phi_{2}}{2})$,
$\sin^{2}(\frac{\phi_{2}}{2})$, 0, $\sin^{2}(\frac{\phi_{3}}{2})$\}.
After Eve's attack, the error probabilities introduced are \{0, 1/2,
1, 1/2\}. The analysis in Base1 can be carried out similarly. So the
QBERs are
\begin{equation}
\begin{aligned}
Base2:
QBER_{2}=\frac{\frac{\sin^{2}(\frac{\phi_{2}}{2})}{2}+\frac{\sin^{2}(\frac{\phi_{3}}{2})}{2}}{\sin^{2}(\frac{\phi_{1}+\phi_{2}}{2})+\sin^{2}(\frac{\phi_{2}}{2})+\sin^{2}(\frac{\phi_{3}}{2})}
\\
Base1:
QBER_{1}=\frac{\frac{\sin^{2}(\frac{\phi_{1}}{2})}{2}+\frac{\sin^{2}(\frac{\phi_{2}}{2})}{2}}{\sin^{2}(\frac{\phi_{2}+\phi_{3}}{2})+\sin^{2}(\frac{\phi_{2}}{2})+\sin^{2}(\frac{\phi_{1}}{2})}
\end{aligned}
\end{equation}

Ref \cite{remapping} assumed $\phi_{1}=\phi_{2}=\phi_{3}=\phi$, then
the total QBER is given by
\begin{equation}
\begin{aligned}
\overline{QBER}=\frac{QBER_{1}+QBER_{2}}{2}=\frac{\sin^{2}(\frac{\phi}{2})}{\sin^{2}(\phi)+2\sin^{2}(\frac{\phi}{2})}
\end{aligned}
\end{equation}

\begin{figure}[hbt]
\centering \resizebox{7.5cm}{!}{\includegraphics{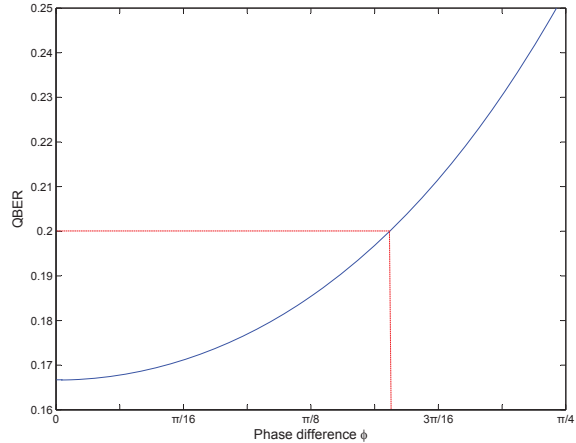}}
\caption{(Color online). QBER of phase-remapping attack. Eve remaps
the four BB84 states with the same new phase difference
($\phi_{1}=\phi_{2}=\phi_{3}=\phi$).} \label{Fig.2}
\end{figure}

As shown in Fig. \ref{Fig.2}, there is a range of ``$\phi$" that
allows QBER to go below 20.0\%, which is tolerable in the BB84
protocol \cite{bb84 proof}. So, if Eve remaps the phase small enough
into this range, she can successfully apply this
``intercept-and-resend" attack.

\section{Experimental phase-remapping attack in a commercial ``plug \& play " QKD system} \label{Exp}
\begin{figure}[hbt]
\centering \resizebox{10cm}{!}{\includegraphics{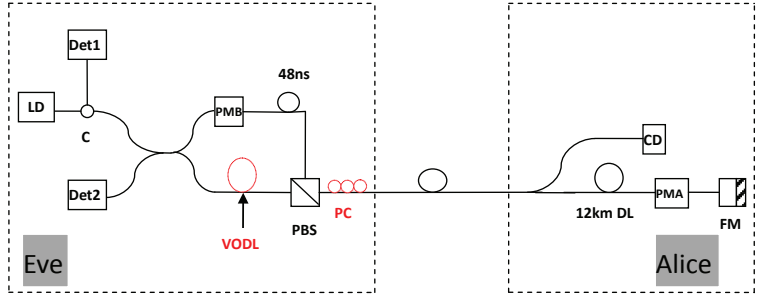}}
\caption{(Color online). Experimental implementation of the
phase-remapping attack in a commercial ID-500 QKD system. Original
QKD system: LD, laser diode. Det1/Det2, single photon detector;
PMA/B, phase modulator; C, circulator. PBS, polarization beam
splitter; CD, classical photodetector; DL, delay line; FM, Faraday
mirror. Our modifications: Eve replaces Bob; VODL, variable optical
delay line; PC, polarization controller.} \label{Fig.3}
\end{figure}

\begin{figure}[hbt]
\centering \resizebox{11cm}{!}{\includegraphics{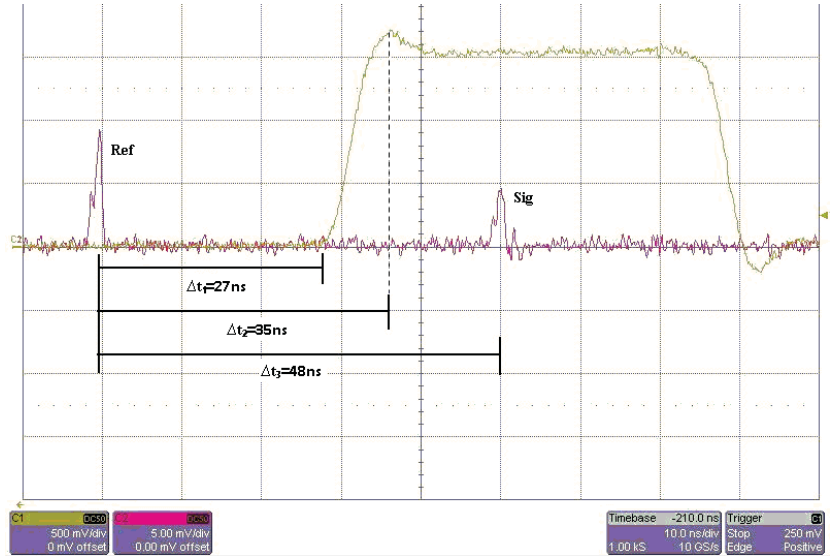}}
\caption{(Color online). Time patterns of the reference pulse (Ref),
the signal pulse (Sig) and the phase modulation signal in the
commercial ID-500 QKD setup. Here, Alice's encoding phase is
\{$\pi$\} and we only show the forward pulses.} \label{Fig.4}
\end{figure}
We implemented the phase remapping attack in a commercial ID-500 QKD
system (manufactured by id Quantique), as shown in Fig. \ref{Fig.3}.
Bob's signal pulse, reference pulse and Alice's phase modulation
signal of the original QKD system are shown in Fig. \ref{Fig.4}.
Note that in Fig. \ref{Fig.4}, since Alice uses the reference pulse
as a trigger signal, the time delay $\Delta t_{1}$ is determined by
the internal delay of Alice's system and can't be controlled by Eve.
On the other hand, since Alice doesn't monitor the arrival time of
the signal pulse, Eve can change the time delay $\Delta t_{3}$
without being detected. Furthermore, the rising edge of the phase
modulation signal is around 8ns, while the width of the laser pulse
is about 500ps. Eve can easily place her pulse on the rising edge to
get partial phase modulation. This specific QKD design opens a
security loophole which allows Eve to launch the phase-remapping
attack.

\begin{figure}[hbt]
\centering \resizebox{8.5cm}{!}{\includegraphics{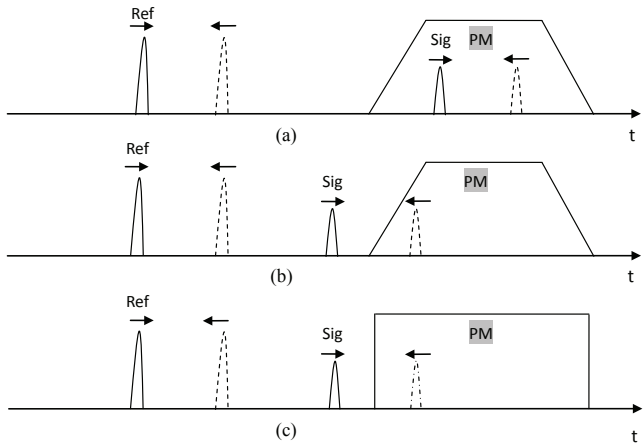}}
\caption{Time pattern of practical phase-remapping attack. Sig:
signal pulse. Ref: reference pulse. PM: phase modulation signal. (a)
Normal QKD operation. (b) Type 1 practical phase-remapping attack.
(c) Type 2 practical phase-remapping attack; here, even if we assume
Alice has a perfect phase modulator with strictly sharp rising and
following edge, Type 2 attack still works.} \label{Fig.5}
\end{figure}

In our experiment, Eve utilized the same setup as Bob to launch her
attack. Eve modified the length of the short arm of her Mach-Zehnder
interferometer by adding a variable optical delay line (VODL in Fig.
\ref{Fig.3}) to shift the time delay between the reference pulse and
the signal pulse. To remap the phase small enough into the low QBER
range, the optimal strategy we found is: by using VODL, Eve shifts
the forward signal pulse out and only the backward signal pulse in
the phase modulation range (see Fig. \ref{Fig.5}(b)).

One important practical challenge in our experiment is polarization
control. The phase modulator in Alice's system is polarization
dependent and has one principle axis. Photons with polarization
aligned with the principle axis will undergo a large phase
modulation, while photons with orthogonal polarization state will
undergo a small phase modulation \cite{Yariv}. In our experiment, we
find the relative modulation magnitude ratio of the two
polarizations is about 1 : 3 \cite{ratio}. In the original ``plug
and play" system, the signal pulse will be modulated twice as it
passes through Alice's phase modulator back and forth (see Fig.
\ref{Fig.5}(a)). Because of the Faraday mirror, the total phase
shift is independent of the polarization state of the signal pulse.
However, since Eve's signal pulse will pass through the modulator at
a different time and be modulated only once (see Fig.
\ref{Fig.5}(b)), the above auto-compensating method will not work.
Eve has to control the polarization direction either aligned with or
orthogonal to the principal axis of the phase modulator when her
signal pulse is modulated. This is achieved by adding a polarization
controller (PC in Fig. \ref{Fig.3}) in Eve's system and adjusting it
carefully. Here, Eve can assume that the polarization has been
aligned properly by maximizing the total counts of D1+D2 (D1 and D2
denote the counts of Det1 and Det2) \cite{confirm}.

By combining variable shifting time and two different polarization
directions, Eve can apply two types of practical phase-remapping
attack:
\begin{enumerate}
\item Type 1 practical attack is shown in Fig. \ref{Fig.5}(b). Eve shifts the forward signal pulse out of the phase modulation signal and the backward
pulse to the rising edge, and adjusts the PC to control the backward
pulse's polarization direction aligned with the modulator's
principal axis. Here, we remark that if the width of laser pulse is
comparable with the rising time of the modulation signal, the type 1
attack will cause an unreasonably high QBER, thus it is easy for
Alice and Bob to detect the attack.
\item Type 2 practical attack is shown in Fig. \ref{Fig.5}(c). Eve shifts the backward
pulse to the plateau region of the phase modulation signal, and
aligns its polarization direction orthogonal to the principal axis.
Since the orthogonal direction has the smallest phase modulation,
Eve can successfully remap the phase small enough into the low QBER
range. One important advantage of type 2 attack is: even if Alice's
phase modulator is good enough with strictly sharp rising and
following edge (force type 1 attack noneffective), Eve can still
apply type 2 attack in practical QKD systems.
\end{enumerate}

\begin{figure}[hbt]
\centering \resizebox{8cm}{!}{\includegraphics{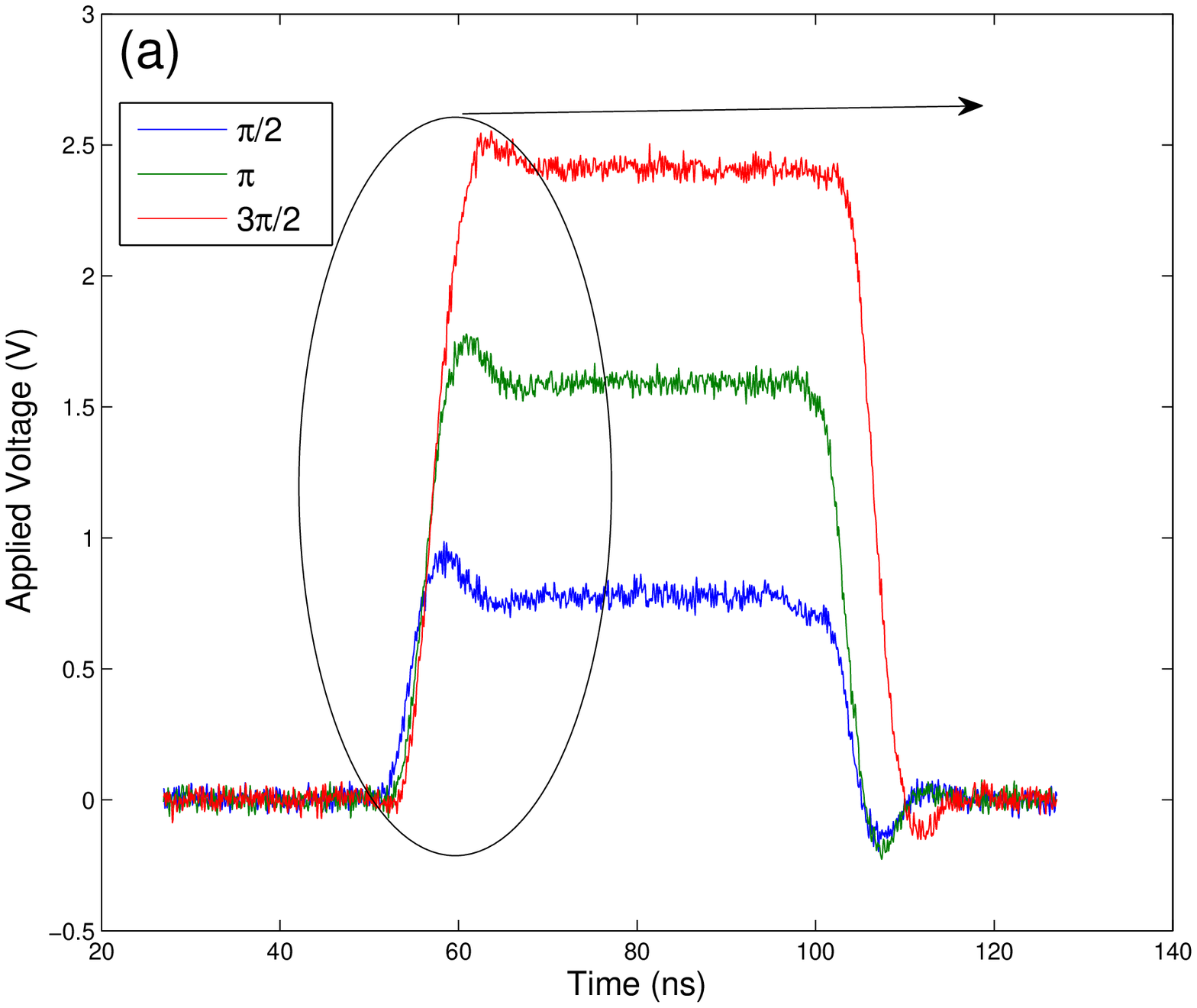}}
\centering \resizebox{7.5cm}{!}{\includegraphics{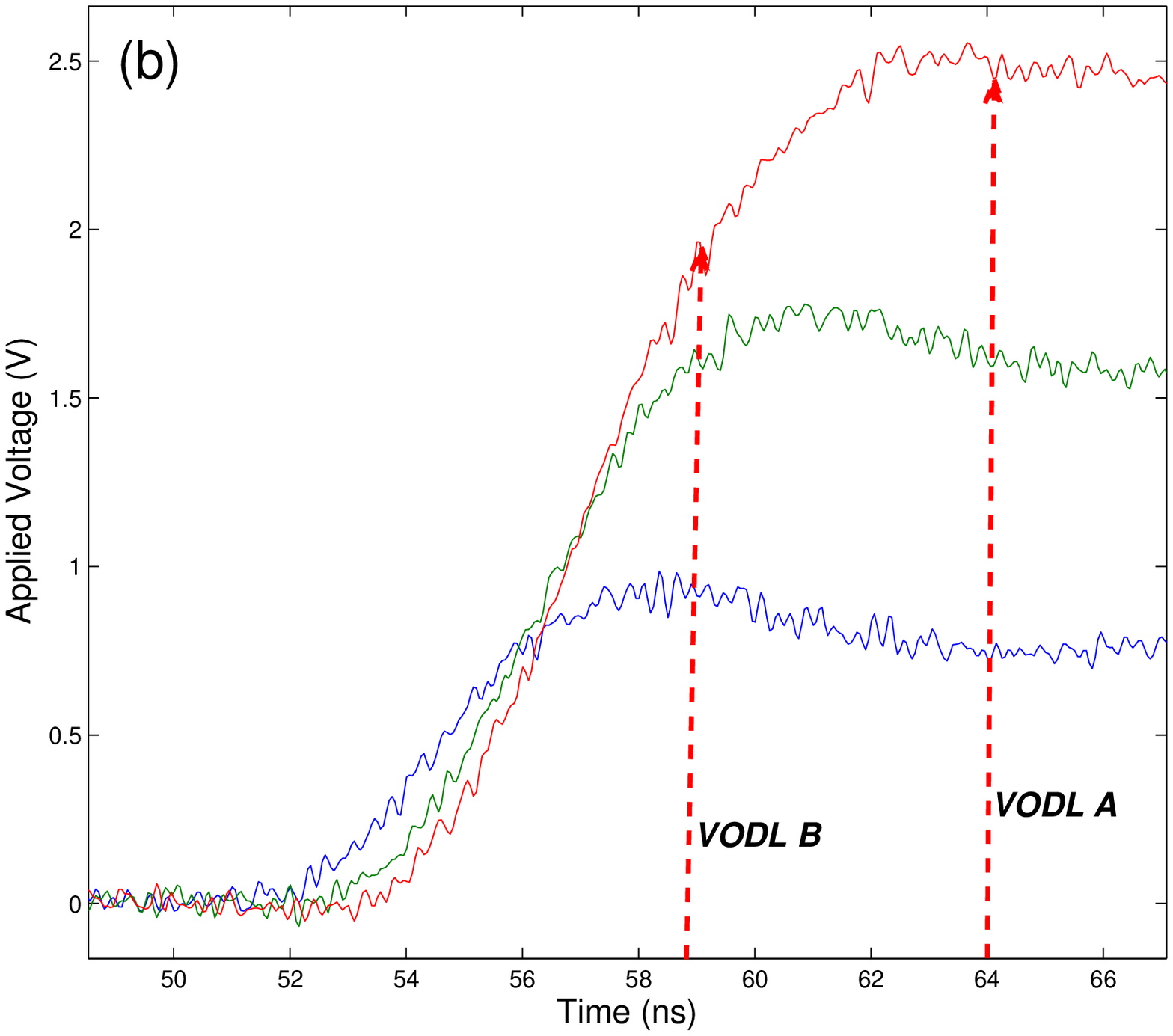}}
\caption{(Color online). (a)Alice's phase modulation signals,
$\pi/2$, $\pi$, and $3\pi/2$, respectively. (b) The zoomed rising
edge of each modulation signal and the approximate time of the
optimal VODL used in our attack.} \label{Fig.6}
\end{figure}

Another challenge is the optimization of VODL. Ref \cite{remapping}
assumed Eve could remap Alice's encoded phase with
$\phi_{1}=\phi_{2}=\phi_{3}$. However, in our experiment, the
relation among  $\phi_{1}$, $\phi_{2}$, and $\phi_{3}$ is more
complicated. As shown in Fig. \ref{Fig.6}, Alice's phase modulation
signals \{$\pi/2$, $\pi$, $3\pi/2$\} not only start at different
times but also have different average rising times \{6.12ns, 7.82ns,
9.47ns\}. Furthermore, there is also an overshoot after the rising
edge, and the time is different from each other. So, if we use
different lengths of VODL to shift the pulse either to the rising
edge or to the overshooting range, the pulse will not undergo
proportional phase modulation. Eve's remapping phase will be
$\phi_{1}\neq\phi_{2}\neq\phi_{3}$. These complicated phases will
thus cause an effect of QBER as shown in Eqn. (1).

In our experiment, the optimal length of VODL was determined by
minimizing the resulting QBER. We finally applied two optimal VODL
(see Fig. \ref{Fig.6}(b)) to launch two types of practical phase
remapping attack: VODL A: 4.65m and VODL B: 5.8m. Our attack
strategy was the one discussed in Section II. Here, we remark two
points: (1) from the time pattern graph in Fig. \ref{Fig.4}, the
laser pulse is narrow enough to allow us to apply type 1 attack; (2)
in type 1 attack, to make the remapping phase small enough, we still
control the polarization of backward pulse orthogonal to the
principal axis.
\section{Experiment results} \label{result}
\begin{table}[!h]\center
\begin{tabular}{c @{\hspace{1cm}} c @{\hspace{1cm}} c @{\hspace{1cm}} c }
\hline $Y_{0}$  & $e_{det}$ &  $\eta_{Bob}$ &  $\mu$\\
\hline$2.11\times10^{-5}$ & $0.38\times10^{-2}$ & $5.82\times10^{-2}$ & 1.39\\
\hline
\end{tabular}
\caption{Experimental parameters.}\label{expresult1}
\end{table}
Some experimental parameters of our ID-500 commercial QKD system,
including dark count rate $Y_{0}$, detector error rate $e_{det}$,
Bob's overall quantum efficiency $\eta_{Bob}$ (including the
detection efficiency of single photon detector), detection
efficiency $\eta$ and mean photon number $\mu$ are listed in Table
\ref{expresult1}. Our transmission distance was a few meters. We
repeated the measurement 10 million times for each state sent by
Alice and the experimental results are shown in Table
\ref{expresult2}.
\begin{table}[!h]\center
\begin{tabular}{c @{\hspace{5.0cm}}c @{\hspace{3.0cm}} c @{\hspace{3.2cm}} c}
& Base0 & Base1 & Base2
\end{tabular}
\begin{tabular}{c @{\hspace{0.5cm}} c @{\hspace{0.5cm}} c @{\hspace{1.5cm}} c @{\hspace{0.5cm}} c @{\hspace{1.5cm}} c  @{\hspace{0.5cm}} c @{\hspace{1.5cm}} c @{\hspace{0.5cm}} c }
\hline State & $\phi_{A}$  & $\phi_{E}$ & D1  & D2 & D1  & D2 & D1  & D2  \\
\hline $0_{1}$ & $0^{\circ}$  & $0^{\circ}$ & 617  & 168910  & 7068  & 174061 & 24841  & 156007 \\
       $0_{2}$ & $90^{\circ}$  & $21.1^{\circ}$ & 5843  & 167206  & 1074  & 179218 & 8557  & 170786  \\
       $1_{1}$ & $180^{\circ}$  & $37.8^{\circ}$ & 18096  & 153962  & 5285  & 174161 & 1239  & 176091 \\
       $1_{2}$ & $270^{\circ}$  & $52.7^{\circ}$ & 33260  & 135616  & 19770  & 160300  & 3530  & 173428 \\
\hline
\end{tabular}
\begin{tabular}{c}
(a)
\end{tabular}

\begin{tabular}{c @{\hspace{5.0cm}}c @{\hspace{3.0cm}} c @{\hspace{3.2cm}} c}
& Base0 & Base1 & Base2
\end{tabular}
\begin{tabular}{c @{\hspace{0.5cm}} c @{\hspace{0.5cm}} c @{\hspace{1.5cm}} c @{\hspace{0.5cm}} c @{\hspace{1.5cm}} c  @{\hspace{0.5cm}} c @{\hspace{1.5cm}} c @{\hspace{0.5cm}} c }
\hline State & $\phi_{A}$  & $\phi_{E}$ & D1  & D2 & D1  & D2 & D1  & D2 \\
\hline $0_{1}$ & $0^{\circ}$  & $0^{\circ}$ & 734  & 171851 & 6617  & 166671  &  16311 & 158479 \\
       $0_{2}$ & $90^{\circ}$  & $23.9^{\circ}$ & 7435  & 165402 & 928  & 169814  &  2772  & 169669  \\
       $1_{1}$ & $180^{\circ}$  & $35.9^{\circ}$ & 16474  & 157385 & 3545  & 166427  & 1348  & 168924 \\
       $1_{2}$ & $270^{\circ}$  & $46.3^{\circ}$ & 26879  & 146917 & 8434  & 161575  & 2672  & 168078 \\
\hline
\end{tabular}
\begin{tabular}{c}
(b)
\end{tabular}
\caption{Experiment results. $\phi_{A}$ is Alice's original standard
BB84 phase. $\phi_{E}$ is the new phase remapped by Eve. D1 (D2) is
the counts number of Det1 (Det2). Here, Eve introduced phase \{0\}
(Base0), \{$\phi_{1}$\} (Base1), and \{$\phi_{1}+\phi_{2}$\}
(Base2), respectively on the reference pulse to measure each state,
and repeated the measurement 10 million times for each state. (a)
Variable Optical Delay Line A (4.65m). (b) Variable Optical Delay
Line B (5.8m).}\label{expresult2}
\end{table}

QBER is analyzed as follows. First, we calculate QBER from the
theoretical model discussed in Section \ref{theory}. The detecting
probability of phase-coding BB84 QKD protocol is
\begin{equation}
\begin{aligned}
Det1:
P_{1}=\frac{1-\cos(\phi_{A}-\phi_{B})}{2}=\sin^{2}(\frac{\phi_{A}-\phi_{B}}{2})=\frac{D1-NY_{0}}{D1+D2-2NY_{0}}\\
Det2:
P_{2}=\frac{1+\cos(\phi_{A}-\phi_{B})}{2}=\cos^{2}(\frac{\phi_{A}-\phi_{B}}{2})=\frac{D2-NY_{0}}{D1+D2-2NY_{0}}
\end{aligned}
\end{equation}
where $N$ denotes the gating number \cite{gate}. Here, we subtract
the dark counts number $NY_{0}$ from each detector's counts number
to get the theoretical detecting probability.

If Eve introduces phase shift \{0\} (Base0) on the reference pulse
to measure each state, the remapping phase $\phi_{E}$ and phase
difference $\phi_{i}$ (i=1,2,3) are
\begin{equation}
\begin{aligned}
\phi_{E}=2\tan^{-1}(\sqrt{\frac{D1-NY_{0}}{D2-NY_{0}}}) \\
\phi_{i}=\phi_{E(i)}-\phi_{E(i-1)}
\end{aligned}
\end{equation}

Using data in Table \ref{expresult2}, from Eqn. (4) and (1), we can
get
\begin{equation}
\begin{aligned}
VODL\ A: \ \phi_{1}=21.1^{\circ}\pm1.1^{\circ} \ \ \ \ \phi_{2}=16.7^{\circ}\pm1.1^{\circ} \ \ \ \ \phi_{3}=14.9^{\circ}\pm1.1^{\circ} \\
QBER_{1(A)}=21\% \ \ \ \  \ \ \ \ QBER_{2(A)}= 13\%  \ \ \ \ \  \ \ \ \  \ \ \ \    \ \ \ \ \ \ \\
VODL\ B: \ \phi_{1}=23.9^{\circ}\pm1.2^{\circ} \ \ \ \ \phi_{2}=12^{\circ}\pm1.2^{\circ} \ \ \ \ \phi_{3}=10.4^{\circ}\pm1.2^{\circ} \\
QBER_{1(B)}=29\% \ \ \ \  \ \ \ \ QBER_{2(B)}= 8\%  \ \ \ \ \  \ \ \
\ \ \ \ \ \ \ \ \  \ \ \ \
\end{aligned}
\end{equation}

The phase error fluctuations are mainly due to the imperfections of
our experimental QKD system. From the results in Table
\ref{expresult2}, we can see that even though Eve uses Base0 to
measure state \{$0_{1}$\}, it still has about ``$600\sim700$" counts
on Det1. These error counts are mostly from the imperfect
interference between the signal pulse and the reference pulse. So,
Eqn. (5) gives the theoretical QBERs introduced by Eve with perfect
detection system.

Now, we calculate QBER via our direct experimental results. From
Table \ref{expresult2}, we can see the total counts (D1+D2) are
almost identical, so Det1's detecting probability for each state is
proportional to D1. The QBERs can be calculated using data in Table
\ref{expresult2}:
\begin{equation}
\begin{aligned}
Base1: QBER_{1}=\frac{\frac{D1_{0_{1}}}{2}+\frac{D1_{1_{1}}}{2}+D1_{0_{2}}}{D1_{0_{1}}+D1_{0_{2}}+D1_{1_{1}}+D1_{1_{2}}}\\
Base2:
QBER_{2}=\frac{\frac{D1_{0_{2}}}{2}+\frac{D1_{1_{2}}}{2}+D1_{1_{1}}}{D1_{0_{1}}+D1_{0_{2}}+D1_{1_{1}}+D1_{1_{2}}}
\end{aligned}
\end{equation}
\begin{equation}
\begin{aligned}
VOL\ A: QBER_{1(A)}=21.8\% \ \ \ \  QBER_{2(A)}= 19.1\%\\
VOL\ B: QBER_{1(B)}=30.8\% \ \ \ \  QBER_{2(B)}= 17.6\%
\end{aligned}
\end{equation}

It is surprising to see that if Eve utilizes the optimal strategy to
combine two types of attack together and carefully chooses the the
probability of each attack to ensure the distribution of bit ``0"
and bit ``1" received by Bob is balanced, the overall QBER will be
\begin{equation}
\begin{aligned}
\overline{QBER}=\frac{QBER_{1(A)}+QBER_{2(B)}}{2}=19.7\%
\end{aligned}
\end{equation}

Note that we used weak coherent pulse (WCP) source in our
experiment. Before calculating the QBERs for single-photon (SP)
source, we emphasize two facts: (1) the phase shift introduced by
the phase modulator is independent of the source. If the source is a
SP, the phase will be also remapped to \{0, $\phi_{1}$,
$\phi_{1}+\phi_{2}$, $\phi_{1}+\phi_{2}+\phi_{3}$\}. (2) Eve's
interference visibility is the same for SP and WCP. Now, assume that
Eve introduces Base1 to launch attack and Det1's detecting
probability for each state is $P_{state}$, i.e. \{$P_{0_{1}}$,
$P_{0_{2}}$, $P_{1_{1}}$, $P_{1_{2}}$\}, Det1's overall gain and
QBERs for the two different sources are given by:
\begin{equation}
\begin{aligned}
SP: \ \ \ \ \ \ \ \ \ \ \ \ \ \ \ \ \ \ \ \ \ \ \ \ \ \ \ \ \ \ \ \ \ \ \ \ \ Q_{sp}=\eta_{Bob}P_{state}+Y_{0} \\
QBER_{sp}=\frac{\eta_{Bob}(\frac{P_{0_{1}}}{2}+\frac{P_{1_{1}}}{2}+P_{0_{2}})+2Y_{0}}{\eta_{Bob}(P_{0_{1}}+P_{0_{2}}+P_{1_{1}}+P_{1_{2}})+4Y_{0}}\\
\end{aligned}
\end{equation}
\begin{equation}
\begin{aligned}
WCP: \ \ Q_{wcp}=\sum_{i=0}^{\infty}(Y_{0}+1-(1-\eta_{Bob}P_{state})^{i})\frac{\mu^{i}}{i!}e^{-\mu}=(1-e^{-\mu\eta_{Bob}P_{state}})+Y_{0}\\
QBER_{wcp}=\frac{2-\frac{e^{-\mu\eta_{Bob}P_{0_{1}}}}{2}-\frac{e^{-\mu\eta_{Bob}P_{1_{1}}}}{2}-e^{-\mu\eta_{Bob}P_{0_{2}}}+2Y_{0}}{4-e^{-\mu\eta_{Bob}P_{0_{1}}}-e^{-\mu\eta_{Bob}P_{0_{2}}}-e^{-\mu\eta_{Bob}P_{1_{1}}}-e^{-\mu\eta_{Bob}P_{1_{2}}}+4Y_{0}}\\
\end{aligned}
\end{equation}

Using Eqn. (9-10) and data in Table \ref{expresult1} and
\ref{expresult2}, the overall QBER difference between SP and WCP for
Eve's optimal strategy (combine two types of attack as Eqn. (8)) is:
\begin{equation}
\begin{aligned}
\Delta\overline{QBER}=\overline{QBER_{sp}}-\overline{QBER_{wcp}}=0.1\%
\end{aligned}
\end{equation}

Therefore, in a practical SP BB84 QKD system, we can expect the QBER
is $\overline{QBER_{sp}}$=19.8\%, which is lower than the tolerate
security bound of 20.0\% \cite{bb84 proof}. The security of SP BB84
QKD system is compromised. Eve can also combine the phase-remapping
attack with a faked state attack together to substantially enhance
its power \cite{remapping}. Finally, we remark that even if we have
only broken a practical SP BB84 QKD system, the security proofs of
both WCP and SP QKD are based on the same assumptions. One key
assumption (Alice prepares her states correctly) has been violated
in our experimental demonstration. So the security proofs can not be
directly applied to a practical QKD system.

\section{Conclusion} \label{con}
We conclude this paper with some general comments. First, let us
consider countermeasures. In the ``plug-and-play'' QKD system, one
specific countermeasure is the following: Alice carefully checks the
arrival time of the reference pulse and the signal pulse by
monitoring with her classical detector (CD in Fig. \ref{Fig.3}).
From the time delay between the two pulses, she can find whether the
time difference has been shifted by Eve, and thus counter Eve's
attack. Moreover, in our attack, Eve only sends two states to Bob.
Alice and Bob can detect this attack by estimating the statistics of
the four BB84 states. Note that, once a security loophole has been
found, it is often easy to develop countermeasures. However, the
unanticipated attacks are the most fatal ones.

Second, this paper mainly focuses on one key assumption in
unconditional security proofs, i.e. Alice prepares the required
states correctly. From a simple experimental demonstration, we show
this assumption can be violated by our attack. So, we emphasize
that, in a practical QKD system, Alice needs to experimentally
verify she is applying the correct modulations on her states. One
possible way in a general QKD system is: after encoding her random
bits, Alice uses a beam splitter to split part of each strong
modulated signal, and then use a $classical$ detector, such as a
power meter (rather than a single-photon detector), to implement a
local measurement to directly verify whether she has performed the
correct modulation. In order to achieve unconditional security with
a practical QKD system, it is useful to perform such a verification
experimentally. In the long term, it is important to work towards
QKD with $testable$ assumptions.

Third, Eve can also maximize her ability to eavesdrop by combining
various attacks. For instance, she may combine the phase-remapping
attack with the time-shift attack to exploit both the imperfections
of Alice's encoding system and Bob's detection system. If she does
so, the QBER might be reduced further. We remark that, it is
impossible to remove all imperfections completely in practice.
Instead of removing them, what we can do is to quantify them
carefully. Once quantified, those imperfections may be taken care of
in security proofs \cite{gllp, individual}. As an example, mismatch
in detection efficiency has been taken into account in the security
proof of \cite{mismatch}.

Finally, our demonstration of the phase remapping attack was done on
a specific implementation of QKD. Notice, however that the
implementation is the one widely used in commercial QKD systems
\cite{commercial}. Moreover, in a general class of QKD systems and
protocols, our work highlights the significance for Alice to verify
the correctness of her preparation. In practice, any QKD is done on
a specific implementation and it has imperfections. If we can't
trust a specific implementation of QKD, one should never use QKD in
the first place.

In summary, we have experimentally demonstrated a technologically
feasible attack, where Eve can get full information and only
introduces a QBER of 19.7\%. A simple ``intercept-and-resend" attack
will normally cause a QBER of 25\% in BB84. So, our result shows
clearly an imperfection in the QKD implementation. The security of a
single photon BB84 QKD system has been compromised. Specially, this
is the first successful ``intercept-and-resend" attack on top of a
commercial bidirectional QKD system. The success of our attack
highlights not only the importance for Alice to verify that she is
encoding the right state during the encoding process, but also, more
generally, the importance of verification of the correctness of each
step of an implementation of a QKD protocol in a practical QKD
system.


\section{Acknowledgments}
We are thankful for the preliminary work of Felipe Corredor, Eva
Markowski, and Chi-Hang Fred Fung. Support of the funding agencies
CFI, CIPI, the CRC program, CIFAR, MITACS, NSERC, OIT, and
QuantumWorks is gratefully acknowledged.



\begin{thebibliography}{00}
\bibitem{det1}
After the Mach-Zehnder interferometer, if the phase difference
between reference and signal pulse is $\pi$ (0), detector1
(detector2) clicks.
\bibitem{ratio}
The relative magnitude ratio is experimentally tested by appling
different voltages on Alice's phase modulator (PMA in Fig.3) to
modulate the signal pulses with the two polarization directions
(adjusted by PC in Fig.3). From the data of applied voltages and
modulating phases, we got the relative ratio is about 1 : 3. Ref.
\cite{Yariv} gives the parameters of LiNbO3 phase modulator and the
relations between phase modulation and the parameters. The relative
ratio is 9.6 : 30.9 (see Section 9.2 and Table 9.2 of Ref
\cite{Yariv}).
\bibitem{confirm}
If the polarization is not properly controlled by PC, after Alice's
modualtion, the original linear polarization state of the signal
pulse will change to circular or ellipse polarization state. So,
when the signal pulse returns back and passes through Eve's PBS (see
Fig.3), part of it will wrongly go to the long arm instead of the
short arm. Since the Detector (Det in Fig.3) is gated, this part
will hit the Detector at a wrong time and thus cannot be detected.
\bibitem{gate}
Here, we repeated the measurement 10 million times for each state.
Notice that, in order to reduce the after-pulsing probability, an
external dead time has been introduced to both detectors after the
detection of a photon by a detector. On average, after each
detection event, the following around 46 gating signals will be
blocked. So, the total gating number $N$ can be estimated by
\begin{equation}
\begin{aligned}
N\approx10^{7}-(D_{1}+D_{2})\times46\approx2.1\times10^{6}
\end{aligned}
\end{equation}
\end{thebibliography}


\end{document}